\documentclass[conference]{IEEEtran}
\title{On Identifying Adversarial Intent Injection in AI-Native 6G Networks}
\usepackage{amsmath}
\usepackage{graphicx}
\usepackage{array}
\usepackage{subcaption}
\usepackage{amssymb, graphicx, booktabs}
\usepackage{algorithm, algorithmic}
\usepackage{comment}
\usepackage{array}
\usepackage{enumitem}       
\usepackage[hyphens]{url}
\usepackage{booktabs}
\usepackage{multirow}
\usepackage[table]{xcolor}
\usepackage{balance}
\usepackage[normalem]{ulem}

\newcolumntype{L}[1]{>{\raggedright\ttfamily\arraybackslash}p{#1}}

\begin{document}

\author{
\IEEEauthorblockN{Nilesh Charkraborty$^1$, Petar Djukic$^2$, Burak Kantarci$^1$}\\
\IEEEauthorblockA{\textit{$^1$University of Ottawa, Ottawa, ON, Canada}\\
\textit{$^2$Nokia Bell Labs, 600 March Road,
Kanata, ON K2K 2E6, Canada}\\
$^1$\{nchakrab, burak.kantarci\}@uottawa.ca,~$^2$petar.djukic@nokia-bell-labs.com}
\vspace{-0.2in}}

\maketitle

\maketitle

\thispagestyle{empty}
\pagestyle{empty}
\begin{abstract}
AI-native 6G networks have brought Intent-Based Networking (IBN) to the forefront, enabling high-level goals to be translated into network configurations. However, this abstraction opens new attack surfaces, primarily adversarial intent injection, where malicious policies are disguised within benign intent flows. The detection of attack instances might become significantly more difficult if the adversaries adopt a stealthy mode of malicious intent injection. With all these in mind, we first define a fine-grained threat model that facilitates the threat of malicious intent injection in an AI-native network. Alongside, we investigate four malicious intent injection strategies$-$ stealth-mode, random distribution, increasing frequency, and decreasing frequency$-$ and propose a dual-path detection framework: (i) a CNN using TF-IDF features for supervised malicious intent detection, and (ii) an AutoEncoder trained exclusively on benign data for one-class malicious intent detection. Our evaluation demonstrates strong detection performance, with accuracy improving to $0.97$ ($\approx 9\%$ gain) and F1-score to $0.98$ ($\approx 36\%$ gain) over the state-of-the-art baseline.\\

\textit{Index Terms$-$} AI-Native Networks, Intent, Adversarial Intent Injection, Threat Detection, Network Security
\end{abstract}
\section{Introduction}
\label{sec:intro}
With the rapid evolution of next-generation networks, the network landscape is shifting towards higher levels of automation, adaptability, and intelligence~\cite{tu2025intent}. Among the emerging paradigms, Intent-Based Networking (IBN) has attracted significant attention for its ability to translate high-level declarative intents into complex configurations~\cite{irtf-nmrg-ibn-usecases-00, globecom2022intent}. 
Intents are typically represented in machine-readable formats such as JSON structures or service templates like TOSCA. In addition, several industry initiatives have proposed domain-specific intent languages, including the TM Forum Intent Ontology and related frameworks developed within telecom standardization bodies such as GSMA. By leveraging artificial intelligence and natural language processing (NLP), IBN enables network operators to specify desired outcomes without dealing with low-level implementation details. The system interprets, assembles, and translates these intents into concrete network policies, thereby enabling streamlined management, improved agility, and reduced operational overhead~\cite{Jacobs2021HeyLumi, leivadeas2022survey}.
However, this abstraction also introduces new security challenges. As IBN frameworks increasingly rely on automation and semantic parsing, they become exposed to a new class of risks-particularly adversarial or unauthorized intent injections, where malicious configurations closely mimic legitimate ones, undermining system integrity~\cite{de2024novel, uchechi2025secured}. 
For example, in~\cite{kim2023intender} the authors consider a scenario in which a legitimate application installs an intent to establish connectivity between two hosts. An adversary with access to the intent interface may submit a second intent that mimics the original policy but with a different identifier or priority. When the controller processes this new intent, it may overwrite or interfere with the flow rules associated with the legitimate one. By subsequently withdrawing the malicious intent, the attacker can silently remove the corresponding forwarding rules, leaving the original intent logically present in the controller but functionally ineffective in the data plane. 
Therefore, such threats, if undetected, can lead to undesirable behaviors ranging from misconfigurations to sophisticated cyberattacks. These concerns highlight the importance of building robust IBN systems that are not only intelligent and efficient but also, resilient to manipulation and adversarial exploitation.

In this paper, we consider a threat model where attackers inject malicious intents into the benign flow of intent traffic using varying frequency patterns- such as stealth-mode injection or gradual increases in malicious activity. These strategies rely on subtle manipulations of contextual dependencies across sequences of intents, where malicious behaviors emerge from the distribution and timing behaviour~\cite{ref-timeBasedThreat} of injected intents.
To the best of our knowledge, this is the first work to explore the possibility of such threats by considering the distribution patterns of malicious intents as contextual cues, and to detect adversarial attempts from the flow of intent traffic. Based on this primary motivation, the contributions of this paper are as follows.

\begin{itemize}[label={}, leftmargin=0pt]
\item\textit{Contribution}~1: We propose a comprehensive adversarial threat model targeting the intent acquisition stage of the IBN pipeline. To emulate realistic attack scenarios, we generate a diverse intent dataset capturing four malicious injection strategies: stealth, random, decreasing frequency, and increasing frequency, where frequency is explicitly modeled as a practical threat dimension~\cite{zhang2021spatio}.

\item\textit{Contribution}~2: We introduce a dual-path detection framework that leverages TF-IDF~\cite{nam2024log}-based features with (i) a supervised CNN classifier and (ii) a one-class AutoEncoder trained exclusively on benign intent sequences. Both models employ a sliding window-based temporal representation to segment intent streams into context-aware windows, enabling the detection of localized and temporally correlated adversarial manipulations.
\end{itemize}
Despite being trained exclusively on benign data, the AutoEncoder achieves strong performance, with accuracy and F$_1$-score exceeding $0.85$ in most scenarios, indicating its effectiveness in detecting previously unseen threat patterns~\cite{zero-day-attack}. The supervised CNN further enhances performance, achieving accuracy and F$1$-scores above $0.95$ in several cases.

The remainder of the paper is organized as follows. Section~\ref{sec:background} establishes the proposed threat model and reviews the related work for gap identification.  Section~\ref{sec:proposed_detection_method} details the construction of the threat detection models, followed by their performance evaluation in Section~\ref{sec:performance}. Finally, Section~\ref{sec:conclusion} presents concluding remarks.

\section{Background and Related Work}
\label{sec:background}
This section first unveils the proposed threat model targeting the intent acquisition stage of the IBN pipeline. Following this, we examine state-of-the-art methods closely related to our work to identify any existing gaps.

\subsection{Threat Model}
\label{subsec:threatmodel}

\begin{figure}[!h]
    \centering
    \includegraphics[width=0.8\linewidth]{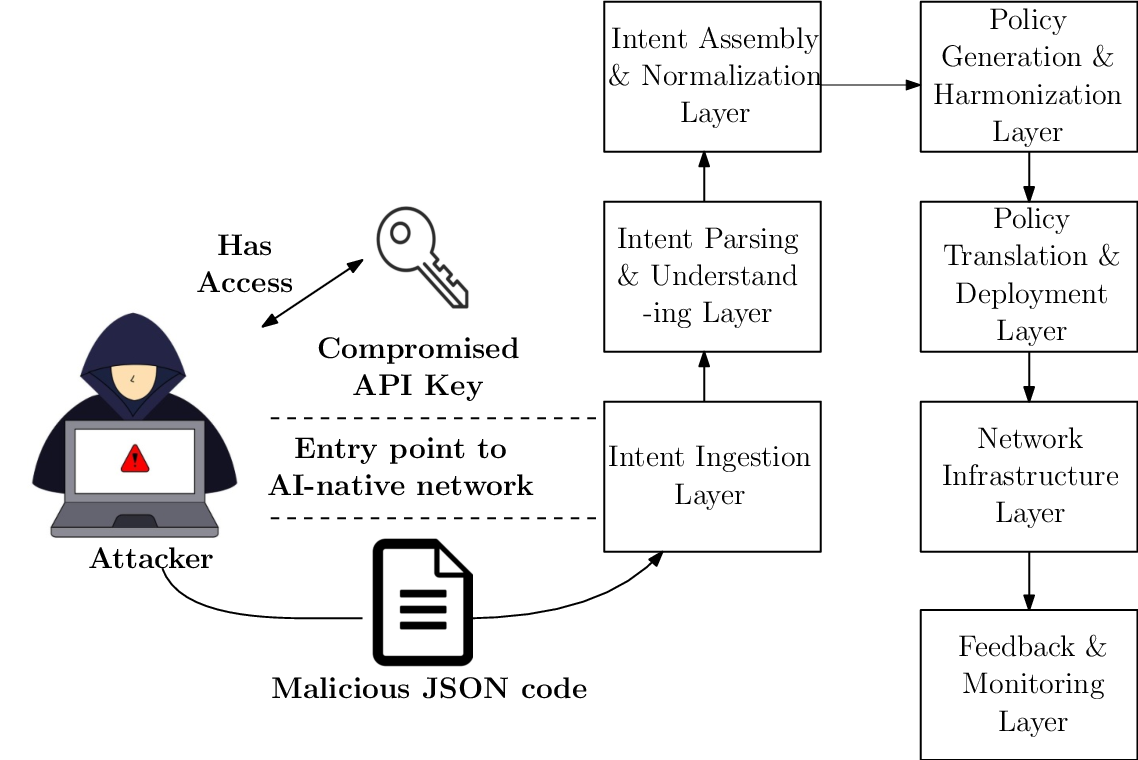}
    \caption{Threat model$-$Malicious intent injection through vulnerable API}
    \label{fig:theme_attack}
\end{figure}

In IBN systems, where intents are expressed in formats such as JSON or TOSCA, an adversary can inject malicious intents into benign flows by exploiting the intent ingestion layer or vulnerable APIs~\cite{de2024novel}. By crafting configurations that closely resemble legitimate requests, these malicious intents evade detection. With the increasing prevalence of API-related threats~\cite{api-threat}, compromised APIs significantly amplify the risk of unauthorized policy injection. Fig.~\ref{fig:theme_attack} illustrates this scenario, where an attacker uses a compromised API key to inject intents that may cause denial of service, privilege escalation~\cite{pecka2022privilege}, traffic redirection~\cite{feng2022off}, or persistent backdoors~\cite{liu2024beyond}, while appearing as routine updates.

We consider four injection strategies with distinct temporal characteristics: (i) stealth injection, where malicious intents follow a Poisson arrival process with a fixed rate; (ii) increasing-frequency injection, where the rate of malicious intents grows over time; (iii) decreasing-frequency injection, where the rate gradually declines; and (iv) random injection, where malicious intents are uniformly distributed across the stream without a predefined arrival model. These strategies capture diverse adversarial behaviors and introduce a temporal dimension to intent injection.

\subsection{Related work}
\label{subsec:relatedwork}
IBN has started gaining traction as a promising paradigm for automating and simplifying network management by translating high-level, goal-oriented intents into low-level configurations~\cite{globecom2022intent-workshop}. Although most early research in IBN has focused on intent formulation, orchestration, and operational efficiency~\cite{Jacobs2021HeyLumi, intent-5G}, to date, only limited work has explored the associated security risks arising from intent-driven network automation. For example, Bringhenti et al.~\cite{Bringhenti2023Survey} present a comprehensive survey of automation challenges in network security, emphasizing the need for context-aware validation mechanisms. Kim et al.~\cite{kim2024security} highlight that the separation between high-level intents and low-level configurations in IBN introduces a semantic gap that can be exploited by adversaries. Trizio et al.~\cite{de2024novel} propose a rule-based malicious intent detection approach for enterprise networks, where \textit{out-of-scope} intents are classified as malicious. However, this work primarily focuses on static rule-level analysis and does not consider temporal variations or stealthy injection strategies, thereby limiting its ability to capture contextual information. Phantom Link attacks further highlight temporal vulnerabilities, where adversaries exploit flow installation delays without altering or manipulating the intents themselves~\cite{weintraub2024exploiting}.

To the best of our knowledge, this work is among the first to emphasize context-based malicious intent detection by varying injection strategies across datasets. Although the intent content remains fixed, the positional context$-$the sequence and spacing of malicious samples$-$captures adversarial behaviors such as stealthy probing and bursty attacks~\cite{kulkarni2025temporal}. Overall, our approach models, simulates, and detects adversarial intent injections with diverse temporal profiles, exposing an underexplored vulnerability and informing potential detection mechanisms~\cite{wu2022ai}.

\section{Methodology}
\label{sec:proposed_detection_method}

We propose a context-aware detection framework for identifying malicious intent injections in IBN by analyzing sequences of intents and their distribution patterns, capturing adversarial behaviors not observable at the individual intent level. TF-IDF-based feature extraction is used as the foundation for two complementary models:

\begin{itemize}[leftmargin=*]
    \item CNN-based classification (supervised)
    \item AutoEncoder-based reconstruction (one-class)
\end{itemize}

Compared to dense embeddings such as Word2Vec, which require large corpora and may blur discrete policy actions (e.g., treating ALLOW, BYPASS, and ENFORCE as semantically similar despite distinct operational meanings), TF-IDF preserves term distinctiveness, enabling more discriminative modeling of intent features.

A 1D CNN is employed due to the sliding-window design with limited context, where detecting short-range dependencies is critical. In contrast, models such as LSTMs target longer temporal dependencies, introduce higher parameter complexity, and are more prone to overfitting under limited data. CNNs, through weight sharing and architectural simplicity, provide an efficient and well-generalizing alternative.

To support detection under limited attack knowledge, a Conv1D AutoEncoder is trained on TF-IDF sequences using sliding windows to learn the temporal structure of benign intent flows. During inference, deviations caused by malicious injection patterns (e.g., stealthy or bursty behavior) are identified via reconstruction error. Unlike traditional one-class methods such as Isolation Forest, One-Class SVM, or k-means, which operate on independent samples, the proposed approach captures localized sequential dependencies, improving sensitivity to subtle temporal anomalies while maintaining efficiency and generalizability.

\subsection{Supervised Model}
\label{subsec:proposed-1}
The intent dataset, consisting of text-based descriptions with each record labeled malicious (1) or safe (0), serves as the input for the supervised algorithm, which follows the key steps outlined below.

\begin{itemize}[label={}, leftmargin=*]
    \item \textit{Step}~1: The intents are converted into numerical vectors using TF-IDF vectorization, with a vocabulary size limited to $500$ features. This captures term-level importance across the corpus.
    \item \textit{Step}~2: To incorporate contextual information, a sliding window of size six is applied across the TF-IDF matrix to generate input sequences. Each sequence is labeled as malicious if any rule within the window is malicious, following a max-label logic.
    \item \textit{Step}~3: The sequences are split into training (75\%) and testing (25\%) sets using a fixed random seed (41).
    \item \textit{Step}~4: To mitigate label imbalance, we compute class weights (W$_c$) by employing the following equation and apply them during training to give more emphasis to the minority class as 
    $W_c=n_s/(n_c\times n_{t_c})$      where \( n_{\text{s}} \) denotes the total number of training samples, \( n_{\text{c}} \) represents the number of unique classes and \( n_{tc} \) denotes number of training samples belonging to the class c.

    \item \textit{Step}~5: 1D CNN architecture is optimized for the detection of temporal patterns in short intent sequences. 
    \item \textit{Step}~6: To further address class imbalance and focus on hard-to-classify samples, we use focal loss with the following formulation: $L(y_{\text{true}}, y_{\text{pred}}) = -\alpha (1 - p_t)^\gamma \log(p_t)$
   where \( p_t \) is the predicted probability for the true class, \( \alpha \) is the weighting factor to balance classes, and \( \gamma \) is the focusing parameter that down-weights easy examples.
  \item \textit{Step}~7: The CNN is trained using the Adam optimizer and the defined focal loss. 
  \item \textit{Step}~8: During inference, an elevated decision threshold is applied to reduce false positives. The model is evaluated using standard performance metrics, including accuracy, precision, recall, and F1-score.
\end{itemize}

\subsection{One-class learning Model}
\label{subsec:proposed-2}
This model follows an encoder-decoder architecture and repeats \textit{Step}~1 to Step~3 from Section~\ref{subsec:proposed-1}. The following three functional features can represent the remaining workflow of this model.

\begin{itemize}[label={}, leftmargin=*]
\item \textit{Encoder:} The encoder employs one-dimensional convolutions to extract localized semantic-temporal patterns from intent sequences, followed by normalization and pooling to stabilize training and reduce dimensionality. A fully connected layer projects the result into a compact latent representation capturing the most salient features.
\item \textit{Decoder:} The decoder reconstructs the original input from the latent representation using fully connected layers, reshaping the output to the original sequence format with a bounded activation to preserve normalization.
\item \textit{Reconstruction loss:} During inference, the AutoEncoder computes the mean absolute reconstruction error for each intent window. A detection threshold is set using a high-percentile reconstruction loss from benign windows; windows exceeding this threshold are flagged as anomalous, indicating potential malicious intent injection.
\end{itemize}

\section{Performance Evaluation}
\label{sec:performance}
Alongside a detailed description of the intent dataset, this section presents the results of our experiments conducted on it. We also compare our findings with those reported in~\cite{de2024novel}, which, to the best of our knowledge, is the only existing work closely related to our contribution. We begin by highlighting the key characteristics and structure of the developed dataset.

\begin{figure}[]
    \centering
    \begin{subfigure}[b]{0.4\textwidth}
        \centering
        \includegraphics[width=\textwidth]{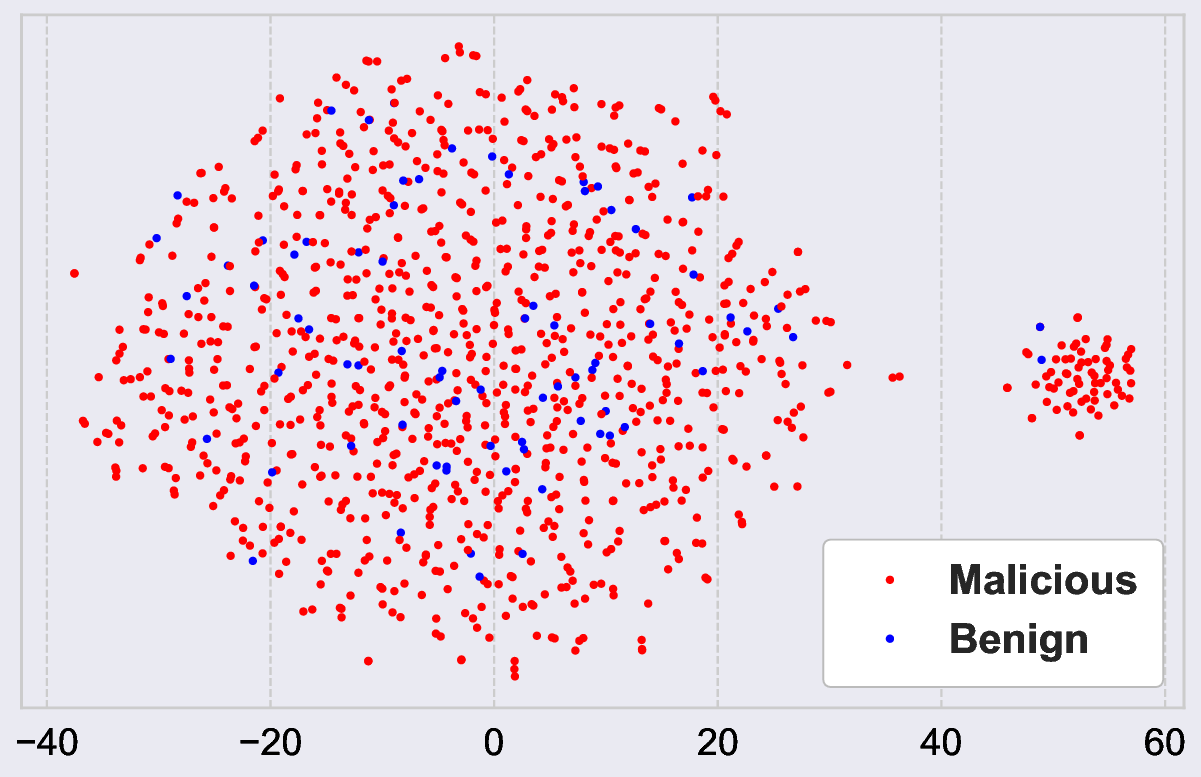}
        \caption{Stealth $\langle$\#B = 76; \#M = 1019$\rangle$}
        \label{fig:sub1}
    \end{subfigure}
    \begin{subfigure}[b]{0.4\textwidth}
        \centering
        \includegraphics[width=\textwidth]{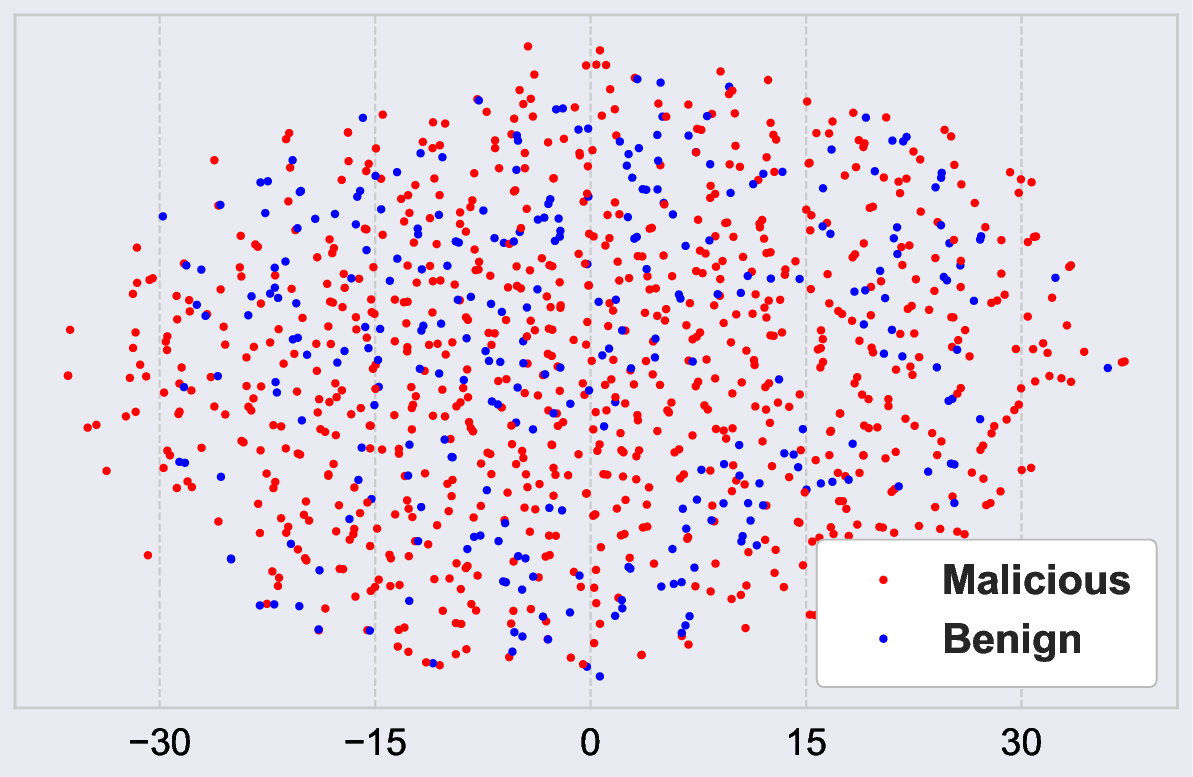}
        \caption{Random $\langle$\#B = 289; \#M = 806$\rangle$}
        \label{fig:sub2}
    \end{subfigure}
    \begin{subfigure}[b]{0.4\textwidth}
        \centering
        \includegraphics[width=\textwidth]{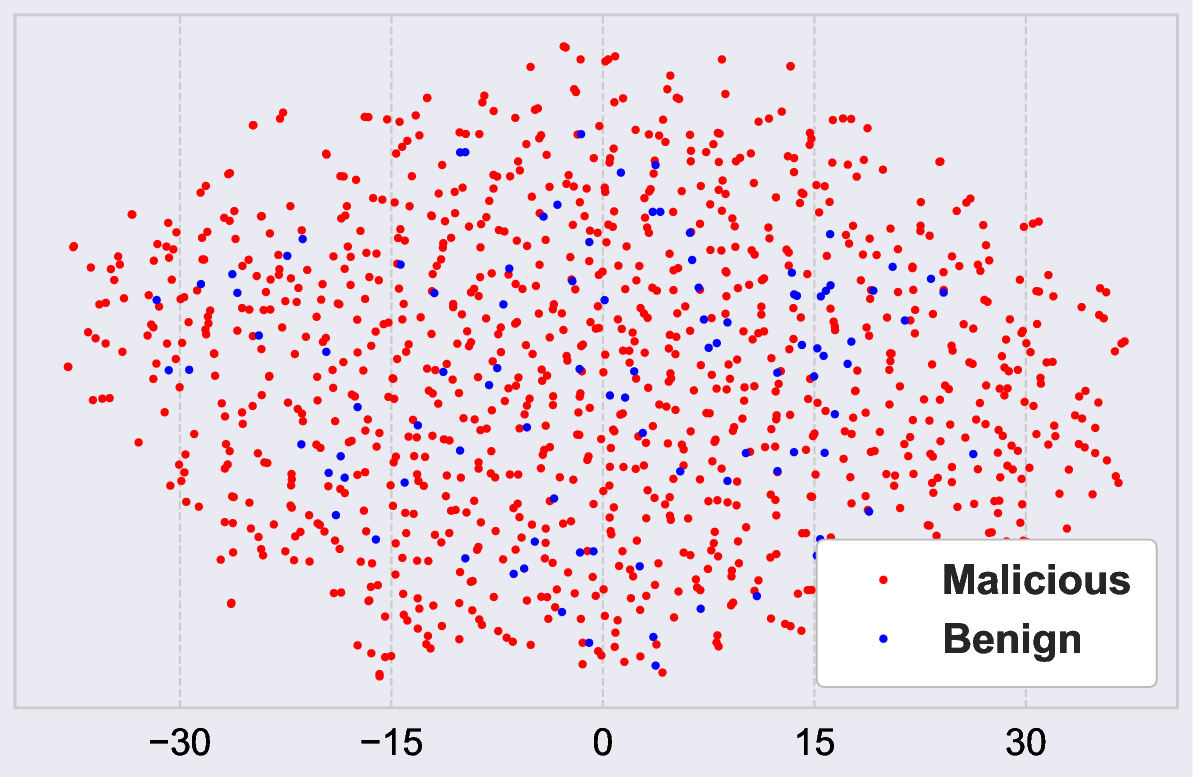}
        \caption{Increasing Frequency $\langle$\#B = 97; \#M = 998$\rangle$}
        \label{fig:sub3}
    \end{subfigure}
    \begin{subfigure}[b]{0.4\textwidth}
        \centering
        \includegraphics[width=\textwidth]{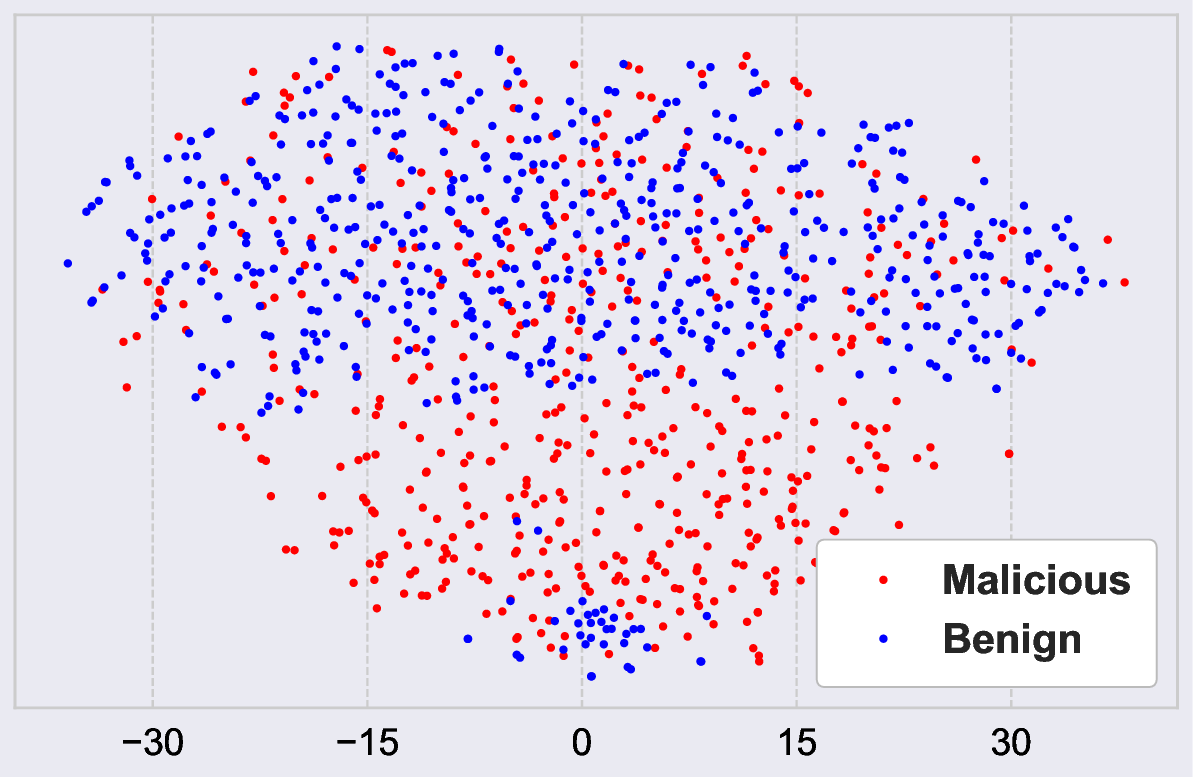}
        \caption{Decreasing Frequency $\langle$\#B = 575; \#M = 520$\rangle$}
        \label{fig:sub4}
    \end{subfigure}
    \caption{t-SNE visualization: For $1095$ windows of window size six, distribution of benign and malicious intents across datasets with underlying temporal dependencies impacting both training and testing phases. \#B and \#M denote the number of benign (blue) and malicious (red) samples, respectively.}
    \label{fig:distribution_malIntent}
\end{figure}

\subsection{Dataset Introduction}
\label{subsec:perform-eval1}
We construct a dataset of 1,100 network intents (250 malicious, 850 benign), partially assisted by a pre-trained LLM, comparable in scale to the proprietary dataset of $755$ intents reported in~\cite{de2024novel}. Malicious intents are generated under four adversarial conditions. Fig.~\ref{fig:distribution_malIntent} illustrates the distribution using t-SNE, where intents are encoded via TF-IDF and projected into two dimensions.
To systematically generate malicious samples, we curate 20 manually verified base intents covering realistic threat scenarios, including DoS, phishing, malware deployment, data exfiltration, and unauthorized access across firewall policies, AI-driven modules, and QoS control. Each base intent is expanded into nine variants that preserve the underlying attack objective while modifying attributes such as action semantics, routing strategy, logging configuration, and endpoint identifiers.
Unlike surface-level linguistic variation, this process introduces protocol- and operation-level transformations, incorporating evasive behaviors such as encrypted tunneling, passive monitoring, delayed or limited logging, and adversarial routing. For example, ACTION:DROP with LOG:ENABLED is transformed into ACTION:NULLROUTE with LOG:LIMITEDMODE, while phishing intents using ROUTING:OVERRIDE are re-expressed via REDIRECTION:ENABLED with INSPECTION:SKIPPED. These transformations emulate realistic obfuscation strategies while preserving attack semantics.
To model contextual ambiguity, we intentionally introduce label noise by relabeling 40 malicious intents as benign and 90 benign intents as malicious based on plausible operational interpretations. This results in a final dataset of 250 malicious and 850 benign intents and forces detection models to rely on semantic reasoning rather than surface patterns.

To ensure the absence of trivial keyword-based cues, we perform a statistical analysis using Chi-square and Fisher’s Exact tests~\cite{agresti2013categorical} over a dictionary of 916 keywords (excluding numerals and stopwords). Keywords are categorized as strong (88), weak (31), or non-discriminative (797) based on statistical significance ($p < 0.05$) and prevalence. A rule-based classifier using only strong discriminators achieves 0.79 accuracy, 0.96 precision, 0.10 recall, and 0.18 F1-score on the full dataset. The low recall confirms that explicit keywords are insufficient for reliable detection, highlighting the need for context-aware approaches.

To characterize malicious intent injection patterns, we model the number of benign intents between consecutive malicious ones as a discrete random variable \(X\). Assuming independent gaps governed by a Poisson process, the probability of observing \(k\) benign samples before the next malicious intent is

\begin{equation}
    P(X = k) = \frac{\lambda^k e^{-\lambda}}{k!}, \quad k = 0, 1, 2, \dots
    \label{eq:poission_eqDistribution}
\end{equation}

For each dataset, we identify malicious intent positions and compute the number of intervening benign samples, whose mean defines the Poisson rate parameter \(\lambda\), representing the expected benign gap between malicious injections. This provides a strategy-specific and interpretable measure of injection spacing, enabling comparative analysis across attack strategies. The estimated rate parameters are found to be \(\lambda \approx 3.40\) (stealth), \(\lambda \approx 3.41\) (random), \(\lambda \approx 1.08\) (decreasing frequency), and \(\lambda \approx 3.01\) (increasing frequency). The estimated values of $\lambda$ indicate that the stealth, random, and increasing-frequency strategies exhibit similar and relatively sparse injection spacing, whereas the decreasing-frequency strategy produces significantly denser malicious injections.

\begin{figure}[!ht]
    \centering
    \begin{subfigure}[b]{0.38\textwidth}
        \includegraphics[width=\textwidth]{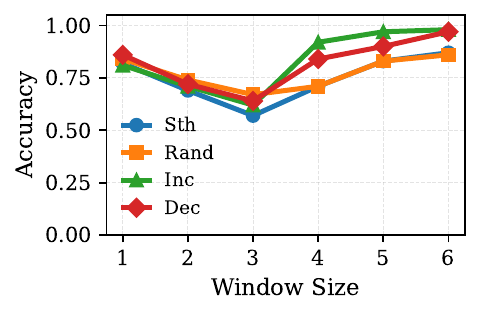}
        \caption{Accuracy (1D CNN)}
    \end{subfigure}
    \begin{subfigure}[b]{0.38\textwidth}
        \includegraphics[width=\textwidth]{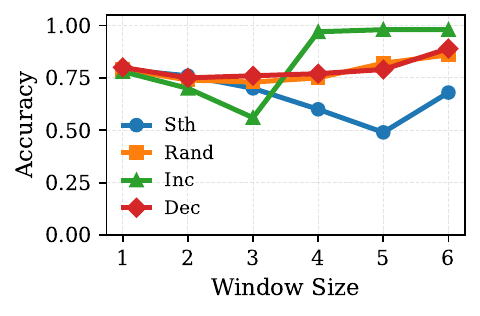}
        \caption{Accuracy (AutoEncoder)}
    \end{subfigure}
    \begin{subfigure}[b]{0.38\textwidth}
        \includegraphics[width=\textwidth]{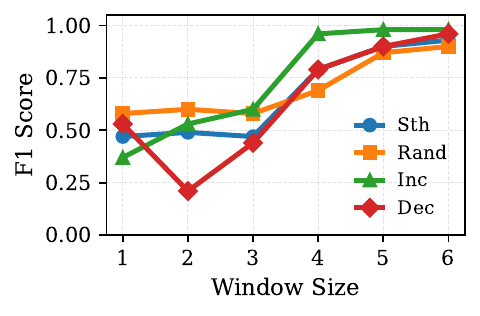}
        \caption{F1-score (1D CNN)}
    \end{subfigure}
    \begin{subfigure}[b]{0.38\textwidth}
        \includegraphics[width=\textwidth]{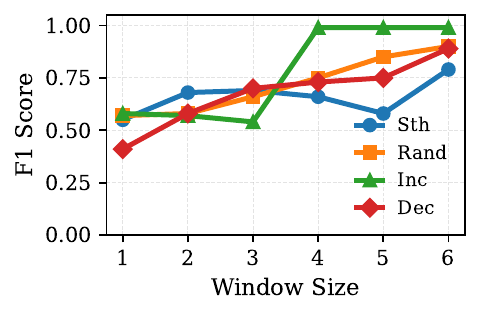}
        \caption{F1-score (AutoEncoder)}
    \end{subfigure}
    \caption{Impact of window size on model performance (accuracy and F1-score) under four malicious intent distributions}
    \label{fig:winSize-performance}
\end{figure}

\subsection{Evaluation Results}
\label{subsec:perform-eval2}

Each intent is encoded using TF-IDF, producing a weighted feature representation that captures both semantic tokens (e.g., \textit{mfa\_auth}, \textit{validate}) and numeric fields (e.g., IP subfields, ports). Table~\ref{tab:tfidf_sample} presents representative examples with their top contributing terms, highlighting recurring discriminative features.
To model short-range dependencies, encoded intents are grouped into sequences using a sliding window of size $1$ to $6$, where a sequence is labeled as malicious if it contains at least one malicious intent. As shown in Fig.~\ref{fig:winSize-performance}, increasing the window size improves accuracy and F1-score for both supervised and one-class models, indicating the importance of temporal context.
For supervised detection, sequences are processed using a 1D CNN with a Conv1D layer (64 filters, kernel size 3), followed by batch normalization, global max pooling, dropout (0.5), and dense layers (16-unit ReLU, sigmoid output). The model is trained using focal loss ($\alpha = 0.25$, $\gamma = 2.0$) with class weighting to address imbalance, for 10 epochs with batch size 16. A prediction threshold of $0.7$ is used to reduce false positives.
For one-class detection, a convolutional autoencoder is trained exclusively on benign sequences. The encoder consists of a Conv1D layer (256 filters, kernel size 3), batch normalization, global max pooling, and a 32-dimensional latent layer, with a mirrored decoder for reconstruction. Training uses MSE loss and Adam optimizer for 10 epochs (batch size 16). During inference, samples are classified as malicious if their reconstruction loss exceeds the 85th percentile of benign losses.

Table~\ref{tab:performance_evaluation-comp} summarizes the results. Under the AutoEncoder, stealth-mode injection yields the lowest recall (0.66), leaving approximately 34\% of malicious samples undetected. For the CNN, random injection is most evasive, with a recall of 0.76. Both models perform best under increasing-frequency injection, achieving recall values of 0.96 (CNN) and 0.98 (AutoEncoder).

\begin{table}[!h]
    \centering
    \renewcommand{\arraystretch}{1.2}
    \caption{Top eight TF-IDF terms and their corresponding weights for each of three representative network intents}
     \begin{tabular}{>{\centering\arraybackslash}m{0.5cm}>{\centering\arraybackslash}m{4cm} >{\centering\arraybackslash}m{2.5cm}}
        \hline\rowcolor{gray!20}
        \textbf{Ind} & \textbf{Intent} & \textbf{TF-IDF Terms (weight)} \\
        \hline
        
        1 & {'type': 'AI-Driven Security', 'action': 'ACTION:VALIDATE; SECURE:MFA\_AUTH; LOG:ENABLED;', 'source': '10.37.197.121', 'destination': '192.168.86.200', 'port': 5000, 'log': True} & 197 (0.387), 121 (0.357), 37 (0.357), 86 (0.354), 200 (0.325), 5000 (0.222), mfa\_auth (0.210), validate (0.210) \\ \hline
    
        2 & {'type': 'QoS Control', 'action': 'ACTION:VALIDATE; SECURE:MFA\_AUTH; LOG:ENABLED;', 'source': '10.74.232.208', 'destination': '192.168.85.202', 'port': 8443, 'log': True} & 74 (0.370), 85 (0.370), 202 (0.363), 208 (0.363), 232 (0.353), mfa\_auth (0.208), validate (0.208), 8443 (0.207) \\ \hline
        
        3 & {'type': 'SLA Compliance', 'action': 'ACTION:MONITOR; AI-SECURITY:ANOMALY DETECTION; LOG:ENABLED;', 'source': '10.177.252.219', 'destination': '192.168.105.153', 'port': 8443, 'log': True} & 105 (0.391), 252 (0.377), 153 (0.352), 177 (0.352), 219 (0.346), 8443 (0.206), anomaly\_detection (0.202), monitor (0.198) \\ \hline
    \end{tabular}
    \label{tab:tfidf_sample}
\end{table}

\subsection{Comparative Results}
We compare the proposed CNN and AutoEncoder models with the DIET classifier proposed by Trizio et al.~\cite{de2024novel}, which is the closest existing baseline for malicious intent detection in IBN. Since the dataset contains a larger proportion of benign samples, accuracy alone may not fully reflect detection effectiveness. Therefore, Table~\ref{tab:performance_evaluation-comp} also reports precision, recall, and F1-score, which provide a more informative evaluation under class imbalance.
As shown in Table~\ref{tab:performance_evaluation-comp}, both proposed models outperform the DIET classifier~\cite{de2024novel}, particularly in recall and F1-score. Although DIET maintains moderate precision across all distributions, its recall remains low (0.49--0.60), indicating that many malicious intents remain undetected.
The proposed CNN consistently achieves strong performance across all four distributions, with the highest F1-score under the stealth and decreasing-frequency settings, and near-best performance under the remaining cases. The AutoEncoder, despite being trained only on benign samples, also performs competitively and achieves the highest recall under the random, increasing-frequency, and decreasing-frequency settings. This suggests that sequence-based anomaly detection can effectively capture deviations from benign intent-flow patterns even without exposure to malicious samples during training.
To summarize, these results highlight the advantage of sequence-aware, context-based modeling over the single-intent classification approach used by DIET~\cite{de2024novel}. 

\begin{table}[t]
\centering
\caption{Performance comparison of the proposed AutoEncoder and CNN models against the DIET classifier under four malicious intent distributions. Accuracy (Acc), Precision (P), Recall (R), and F1-score (F1) are reported. The symbol $\uparrow$ indicates the best-performing method for each metric under a given distribution.}
\label{tab:performance_evaluation-comp}
\setlength{\tabcolsep}{4pt}
\resizebox{0.48\textwidth}{!}{
\begin{tabular}{lcccccccccccc}
\toprule
\multirow{2}{*}{\textbf{Distribution}} &
\multicolumn{4}{c}{\textbf{AutoEncoder}} &
\multicolumn{4}{c}{\textbf{CNN}} &
\multicolumn{4}{c}{\textbf{DIET}~\cite{de2024novel}} \\
\cmidrule(lr){2-5}\cmidrule(lr){6-9}\cmidrule(lr){10-13}
& \textbf{Acc} & \textbf{P} & \textbf{R} & \textbf{F1}
& \textbf{Acc} & \textbf{P} & \textbf{R} & \textbf{F1}
& \textbf{Acc} & \textbf{P} & \textbf{R} & \textbf{F1} \\
\midrule
Stealth    
& 0.68 & 0.98$\uparrow$ & 0.66 & 0.79 
& 0.88 & 0.97 & 0.90$\uparrow$ & 0.93$\uparrow$ 
& 0.88 & 0.84 & 0.54 & 0.65 \\

Random     
& 0.86 & 0.93 & 0.87$\uparrow$ & 0.90$\uparrow$ 
& 0.82 & 0.99$\uparrow$ & 0.76 & 0.86 
& 0.88 & 0.86 & 0.59 & 0.70 \\

Increasing 
& 0.96 & 0.97 & 0.98$\uparrow$ & 0.98$\uparrow$ 
& 0.97$\uparrow$ & 0.99$\uparrow$ & 0.96 & 0.97 
& 0.89 & 0.90 & 0.60 & 0.72 \\

Decreasing 
& 0.89 & 0.83 & 0.96$\uparrow$ & 0.88 
& 0.97$\uparrow$ & 0.98$\uparrow$ & 0.93 & 0.95$\uparrow$ 
& 0.88 & 0.93 & 0.49 & 0.64 \\
\bottomrule
\end{tabular}}
\end{table}

\section{Conclusion and Future Work}
\label{sec:conclusion}
This work addresses adversarial intent injection in AI-native 6G networks by introducing a threat model focused on the intent acquisition stage of the IBN pipeline. We construct a dataset covering four attack distributions and propose a dual-path detection framework combining supervised (CNN) and one-class (AutoEncoder) models. Experimental results show that the CNN achieves consistently strong performance across all scenarios (average recall 0.89, F1-score 0.93), while the AutoEncoder is particularly effective against increasing and decreasing injection strategies (average recall 0.87, F1-score 0.89).
Future work will extend the dataset with more diverse and nuanced JSON-based policy configurations to better reflect real-world IBN environments. We also plan to improve interpretability through explainable AI techniques, enabling greater transparency and trust in detection decisions. These directions aim to support more adaptive and secure IBN frameworks against evolving adversarial strategies in autonomous 6G networks.

\section*{Acknowldgement}
This work is supported in part by Mitacs Accelerate program under project number IT43178, and in part by the Natural Sciences and Engineering Research Council of Canada under the Discovery and CREATE TRAVERSAL programs.

\balance
\bibliographystyle{IEEEtran}
\bibliography{reference}
 
\end{document}